\documentstyle[aps,pra,twocolumn,epsfig,eqsecnum]{revtex}
\begin{document}
\draft
\title
{Quantum-limited linewidth of a good-cavity laser: An analytical 
theory from near to far above threshold}
\author{Ulrike Herzog$^1$ and J\'{a}nos A. Bergou$^2$}
\address{$^1$Institut f\"ur Physik,  Humboldt-Universit\"at zu 
    Berlin, Invalidenstrasse 110, D-10115 Berlin, Germany\\
    $^2$Department of Physics, Hunter College, City
    University of New York, 695 Park Avenue, New York, NY 10021, 
    USA \\
    and Institute of Physics, Janus Pannonius University,
    H-7624 P\'{e}cs, Ifj\'{u}s\'{a}g \'{u}tja 6, Hungary}

\maketitle

\begin{abstract}
 The problem of the quantum-limited or intrinsic linewidth of a
good-cavity laser is revisited. Starting from the Scully-Lamb 
master equation, we present a fully analytical treatment to 
determine the correlation function and the 
spectrum of the cavity field at steady state. 
For this purpose, we develop an analytical approximation 
method that implicitly 
incorporates the microscopic fluctuations of both 
the phase and intensity of the field, and, in addition,
takes full account of the saturation of the nonlinear gain.
Our main result is a simple formula for the quantum-limited
linewidth that is valid from near to far above threshold
and also includes the presence of thermal photons. 
Close to the threshold, the linewidth
is twice as large as predicted by the standard phase-diffusion 
treatment neglecting intensity fluctuations, 
and even 50\% above threshold the increase is still considerable. 
In general, quantum fluctuations of the intensity are present 
and continue to influence the linewidth as long as the
photon-number distribution is not strictly Poissonian. This 
inherent relationship is displayed by a formula relating the 
linewidth and the Mandel Q-parameter.
More than 100\% above treshold the linewidth is found to be
smaller than predicted by the standard treatment, since the
simple phase-diffusion model increasingly overestimates the
rate of phase fluctuations by
neglecting gain saturation. In the limit of a very
large mean photon number the expected perfectly
coherent classical field is obtained.

\end{abstract}

\section{Introduction}

When the radiation field is described classically and spontaneous
emission is neglected in comparison to induced emission, 
the steady-state linewidth of an ideal single-mode laser,
i. e. of a laser with a perfectly stabilized classical
intensity, has the shape of a 
$\delta$-function. This is due to the fact that the cavity losses 
are exactly compensated for by the gain and, as a result, the 
field in the resonator remains constant. 
From a fully quantized description of the electromagnetic 
field, however, it follows  that the laser linewidth cannot be 
smaller than a certain quantum limit, related to the well-known
Schawlow-Townes linewidth, which is inversely proportional to 
the laser intensity \cite{schawlow}.
This limit has been studied extensively in the past 
decades (see Refs. \cite{haken,sargent,sctext,orszag} and references
therein).
In view of the importance of stable coherent signals for various
high-precision measurements and because of an ongoing interest in
fundamental problems of quantum optics, renewed attention has been 
paid recently to the  quantum limitation of the laser linewidth.
The investigations have been extended to cover bad-cavity lasers
\cite{badcavity}, lasers without inversion \cite{fleischh},
and a chaotic laser cavity \cite{patra}.
Different systems have been devised in this context to reduce the 
quantum limit of the linewidth by means of correlated spontaneous 
emission schemes \cite{corr}. Recently even theoretical models
for light amplification without stimulated emission have been 
investigated in order to obtain a reduced ultimate quantum limit 
of the laser linewidth \cite{wiseman}.
                                                  
The quantum-limited laser linewidth is also called
the intrinsic or natural linewidth of the laser.
Its origin, like the origin of the microscopic 
intensity fluctuations, lies in the fact 
that in the stationary regime of operation 
the balance between the gain and loss processes
sustains a constant average field only but, 
due to the quantum nature of these processes, 
fluctuations of the field around its mean are induced on a 
microscopic scale.
We note that the resonator losses are caused by the outcoupling
of the field through the output mirror as well as by any
additional linear damping process such as absorption.
For good-cavity lasers the combined effect of these losses 
can be described by a single cavity-damping
constant $\gamma$ that is the sum of the constants referring to the 
individual processes.
The usual treatment \cite{haken,sargent,sctext,orszag} of the intrinsic
laser linewidth rests on the approximation that the linewidth arises
from fluctuations of the phase of the field described by
phase diffusion. 

In this paper we rely on the Scully-Lamb model
\cite{sargent} of the laser since it is applicable for an arbitrary
strong saturation of the lasing atomic transition. Neglecting intensity
fluctuations, a simple analytic expression for the linewidth has been
derived in this model by means of different methods
\cite{sargent,sctext,orszag}. On the other hand, the intrinsic laser
linewidth can be determined exactly by numerically calculating the
first-order correlation function of the field and performing the Fourier
transform of the latter. Thus both the effects of fluctuations of the
phase of the field and of its amplitude, or intensity, respectively,
are implicitly taken into account. Investigations of this kind have
been performed by N. Lu \cite{lu1} who started from the 
Scully-Lamb master equation \cite{sargent} and found numerically that near
threshold the intrinsic laser linewidth is up to twice as large as
that given by the phase-diffusion coefficient at the mean photon number.

In the present contribution we derive an analytical
expression for the quantum-limited laser linewidth by means of
investigating the two-time first-order correlations of the field.
The treatment is restricted to single-mode
lasers in the good-cavity limit, i. e. we make the usual
assumption that the cavity damping time $\gamma^{-1}$ is
long in comparison to all other relevant time scales. This ensures
that the time dependence of the polarization and population
inversion of the gain medium can be adiabatically eliminated
and the Scully-Lamb model of the laser can be applied.
Our approximation scheme is an extension of an analytical method 
developed previously by one of the authors for calculating
photon-number variances \cite{m,h1}. It makes use of the fact 
that the photon-number distribution of the laser radiation is
strongly peaked at a large mean photon number. Therefore it is not 
necessary to study the density-matrix elements of the field in 
detail but it suffices to directly evaluate the expectation
values and correlation functions of interest,
in the approximation of small fluctuations.

The paper is organized as follows. In Sec. II we develop a
general approximation method for the determination of the
first-order correlation function of the field in the resonator
of a micromaser or laser. The method is applied in Sec. III to
study the laser linewidth. In order to reveal the influence
of intensity fluctuations on the latter, an expression for the
photon number variance of the laser in the
presence of thermal photons is also derived in this section.
Our results are discussed in Sec. IV and compared to the standard
linewidth formula ensuing from the phase diffusion approximation.
Finally a summary is given in Sec. V.

\section{General approximation method}

The power spectrum of a radiation field is defined as the Fourier
transform of its normalized first-order correlation function. 
When the field is a single-mode field with frequency
$\nu$, the steady-state spectrum is given by
\begin{equation}
S(\omega) = \frac{1}{\pi} {\rm Re} \int^{\infty}_0
\frac{\langle a^{\dag}(t) a(0)\rangle_s}
{\langle a^{\dag} a \rangle_s} {\rm e}^{-i(\omega - \nu)t} dt,
\label{1}
\end{equation}
where $a$ and $a^{\dag}$ are the photon annihilation and
creation operators in the interaction picture and the subscript $s$
denotes the stationary state.
We consider a single-mode radiation field
contained in a leaky cavity and being sustained by a gain mechanism.
The overall losses of the field mode are supposed to be due to 
the coupling of the cavity field to the environment, modeled by a 
reservoir in thermal equilibrium.
Therefore the damping of the field can be described by
the standard master equation $\dot{\rho}=L\rho$
for its reduced density operator $\rho$, where
the action of the superoperator $L$ is defined as
\begin{eqnarray}
 L\rho & = & -\frac{\gamma}{2}(1+n_b)
 (a^{\dag} a \rho- 2 a \rho a^{\dag} +\rho a^{\dag}a)\nonumber\\
     & & -\frac{\gamma}{2} n_b
 (a a^{\dag} \rho- 2 a^{\dag} \rho a +\rho a a^{\dag}) .
\label{2}
\end{eqnarray}
Here $\gamma$ is the cavity damping constant, and $n_b$ denotes the
mean number of thermal photons in the cavity.
We also note that for any field operator $\sigma$
the relations,
\begin{eqnarray}
{\rm Tr} (a^{\dag}L\sigma) & = &- \frac{\gamma}{2}
                             {\rm Tr} (a^{\dag}\sigma),
\label{3} \\
{\rm Tr} (a^{\dag}aL\sigma) & = & - \gamma {\rm Tr} (a^{\dag}a\sigma)
                                + \gamma n_b {\rm Tr} \sigma ,
\label{4}
\end{eqnarray}
follow from Eq. (\ref{2}).
To model the gain in a simple way, we assume that excited atoms are 
injected into the cavity with rate $r$. The atoms are supposed
to interact with the field one after the other during a time
$\tau$ that is
negligibly short in comparison to both the cavity damping time
$\gamma^{-1}$
and the mean time interval between successive atoms.
The effect of a single atom on the field can be
formally written as
$\rho(t + \tau) = M \rho(t)$,
where the superoperator $M$ depends on the specific kind
of interaction process. For a micromaser
the interaction time $\tau$ is given by the transit time of the 
atoms
through the microwave cavity, and $M=M(\tau)$ has to be obtained 
in the usual way \cite{microm} from
the Jaynes-Cummings Hamiltonian \cite{jaynes}
for the atom-field interaction
by performing the trace with respect to the atoms.
In order to describe a laser, it is assumed
that excited two-level atoms are injected into a resonant
cavity and interact independently with the field during time
intervals that are
determined by the survival time $\Gamma^{-1}$
 of the atoms as effective two-level
systems, before they decay into different energy states \cite{sctext}.
The superoperator $M$ that has to be used for the laser
is found by averaging $M(\tau)$ with respect to the interaction 
time $\tau$, according to  
$M = \Gamma \int_0^{\infty}M(\tau)
 {\rm e}^{-\Gamma \tau} d\tau$ \cite{sctext}.

  When the injection times of the atoms are uncorrelated, i. e. for
Poissonian pumping, the evolution of the field due to the combined
action of the gain and loss mechanisms  obeys the master equation
\begin{equation}
\dot{\rho}= r(M-1)\rho + L\rho ,
\label{5}
\end{equation}
which has the formal solution $\rho(t) = V(t) \rho(0)$ with
\begin{equation}
V(t) = {\rm exp}\{[r(M-1)+L]t\}.
\label{5a}
\end{equation}

Due to the Markovian character of the master equation
the two-time correlation function necessary to determine the
spectrum can be easily calculated.
For the stationary state we find
\begin{equation}
\langle a^{\dag}(t) a(0) \rangle_s  =
{\rm Tr} [a^{\dag} V(t)(a \overline{\rho})] ,
\label{6}
\end{equation}
with
$\overline{\rho} = \lim_{t\rightarrow \infty} \rho(t)$
denoting the steady-state density operator.
Making use of Eq. (\ref{5a})
and taking into account the relation (\ref{3}),
we obtain from Eq. (\ref{6}) the differential equation
\begin{equation}
\frac{ d}{d t}\langle a^{\dag}(t) a(0) \rangle_s
=\left\{r[b(t)-1] - \frac{\gamma}{2} \right\}
\langle a^{\dag}(t) a(0) \rangle_s ,
\label{7}
\end{equation}
where we have introduced the abbreviation
\begin{equation}
b(t)=
\frac{{\rm Tr}[a^{\dag}M \sigma(t)]}
{\langle a^{\dag}(t) a(0) \rangle_s} =
\frac{\displaystyle{\sum_{n=1}^{\infty} \sqrt{n}[M\sigma(t)]_{n-1,n}}}
{\displaystyle{\sum_{n=1}^{\infty} \sqrt{n}\sigma_{n-1,n}(t)}}
\label{8}
\end{equation}
with
\begin{equation}
\sigma(t)=V(t)(a\overline{\rho}).
\label{8a}
\end{equation}
Under steady-state conditions it is possible
to eliminate the injection rate $r$ from Eq. (\ref{7}) by
expressing it in terms of
field expectation values and the cavity decay rate.
To this end we start from the steady-state relation
\begin{equation}
\frac{d}{dt}\overline{n}
=\sum_{n=0}^{\infty}  n\dot{\overline{\rho}}_{n,n}=
{\rm Tr}[a^{\dag}a\dot{V}\overline{\rho}]=0.
\label{8b}
\end{equation}
By inserting  Eq. (\ref{5a}) and utilizing Eq.(\ref{4})
we arrive at the photon-number balance equation
\begin{equation}
r[{\rm Tr}(a^{\dag}a M\overline\rho) - \overline{n}]
= \gamma (\overline{n}-n_b) ,
\label{9}
\end{equation}
where 
$\overline{n} = \langle a^{\dag} a \rangle_s
={\rm Tr}(a^{\dag}a{\overline\rho})$
is the steady-state mean photon number
$\langle a^{\dag}(0)a(0) \rangle$.

So far all equations hold exactly and can be applied to lasers
as well as to micromasers with Poissonian pumping.
In order to obtain an analytical expression for the spectrum,
it is necessary to calculate the quantity $b$ in Eq. (\ref{7}).
For this purpose we use an approximation method that
rests on the assumption that
the steady-state photon-number distribution in the cavity
is strongly peaked at a large mean photon number $\overline{n}$.
Since the order of magnitude of the width of
the photon-number distribution is determined
by $\sqrt{\Delta n^2}$, we assume that the relations
\begin{equation}
\overline{n} \gg 1, \hspace{1cm}
\sqrt{\Delta n^2}=
(\overline{n^2}-\overline{n}^2)^\frac{1}{2} \ll \overline{n}
\label{10} 
\end{equation}
are fulfilled for the mean photon number and its variance
$\Delta n^2$.
By means of performing an expansion with respect to
suitable parameters in Eq. (\ref{8}) and
replacing $\sigma(t)$ by its initial value $\sigma(0)$
in small terms in this expansion,
it is possible to obtain an approximate expression for $b$.
It will not depend on time, provided that the leading term
in the expansion proves to be time-independent.
In this case the value of $b$ depends only on
$\overline{\rho}$ and on the specific form of $M$, i. e. the
specific kind of the atom-field interaction process,
and we can easily integrate Eq. (\ref{7}) to obtain
\begin{equation}
\langle a^{\dag}(t) a(0) \rangle_s
 =  \overline{n}\, {\rm e}^{[r(b-1)-\frac{\gamma}{2}]t}.
\label{11}
\end{equation}
Fourier transformation according to Eq. (\ref{1})
yields a Lorentzian steady-state spectrum
that is centered at the frequency
$\omega_0=\nu+\frac{r}{2} {\rm Im}\,b$.
The linewidth (full width at half maximum) is
given by
\begin{equation}
\Delta\omega =  \gamma + 2r(1-{\rm Re}\,b).
\label{12}
\end{equation}
As required, the linewidth reduces to the empty-cavity linewidth
$\gamma$ when $r=0$ or when $b=1$, i. e. when either no atoms are 
injected at all, or when the atoms do not interact with the field 
and $M$ is equal to the unit operator.
Under these conditions the steady-state is, of course, the 
thermal state with mean photon number $n_b$, or,
for $n_b = 0$, the vacuum.

We assume that the superoperator $M$ in Eq. (\ref{5})
is due to resonant one-photon interaction of the field 
with a two-level atom, initially in its excited state.
In the photon-number representation, $M$ has the general form
\begin{equation}
(M\rho)_{n,m}= A_{n,m} \rho_{n,m}
                 + B_{n-1,m-1} \rho_{n-1,m-1},
\label{13}
\end{equation}                              
where the coefficients $A_{n,m}$ and $B_{n,m}$ are different for 
the cases of a micromaser or a laser, respectively. 
It then follows that the steady-state
density operator is diagonal in the photon-number representation,
as can be easily shown with the help of Eqs. (\ref{2}) and (\ref{5}).
Therefore the expectation value of the field (which in the 
steady state corresponds to its time average) vanishes,
\begin{equation}
\langle a \rangle_s = {\rm Tr}(a \overline{\rho})=0.
\label{13a}
\end{equation}
In terms of the amplitude and phase of the expectation value of the 
field, $\langle a \rangle_s$, Eq. (\ref{13a}) requires
that for parameter values for which the amplitude is 
fixed, the phase is uniformly distributed between $0$ and 
$2\pi$, i. e. all phase values are equally probable at steady state.
We note that this holds for lasers as well as for micromasers 
unless the atoms are 
injected in a definite superposition of their energy states.

In general, a correlation exists between the values of the 
field at different times that decays with increasing time 
difference. It is this decay that determines the spectrum and the 
linewidth according to Eqs. (\ref{1}) and (\ref{11}).

\section{The laser linewidth}

For the case of the laser, the coefficients
$A_{n,m}$ and $B_{n,m}$ in Eq. (\ref{13})
can be written as \cite{sargent,sctext,orszag}
\begin{equation}
A_{n,m}= 1-
      \frac{\chi\left[1+\frac{1}{2}(n+m)\right]+
      \frac{1}{8}\chi^2 (n-m)^2}
           {2+\chi(n+m+2)+\frac{1}{8}\chi^2 (n-m)^2} ,            
\label{14}
\end{equation}
and
\begin{equation}
B_{n,m}= \frac{\chi\sqrt{(n+1)(m+1)}}
           {2+\chi(n+m+2)+\frac{1}{8}\chi^2 (n-m)^2},
\label{15}
\end{equation}
where we introduced the saturation parameter $\chi=4g^2/\Gamma^2$
with $g$ and $\Gamma^{-1}$ denoting the
atom-field coupling constant and the average lifetime
of the atom as a two-level system, respectively.
After  changing the index of summation
appropriately, from Eqs. (\ref{8}) and (\ref{13}) we obtain
\begin{equation}
b(t)=1+\frac{1}{4} 
\frac{\displaystyle{\sum_{n=1}^{\infty}\frac{\chi}{1+\chi(n+\frac{1}{2})}
\sqrt{n}\sigma_{n-1,n}(t)}}
{\displaystyle\sum_{n=1}^{\infty}\sqrt{n}\sigma_{n-1,n}(t)}.
\label{16}
\end{equation}
Here use has been made of the fact that the terms proportional
to $\chi^2$ in Eqs. (\ref{14}) and (\ref{15}) can be neglected
for $|n-m| = 1$ since under normal conditions the relation
$\chi^2 \ll 1$ \cite{sctext} is fulfilled.                        
To evaluate $b$, we now apply the approximation
\begin{equation}
\frac{1}{1+\chi(n+\frac{1}{2})} \approx
\frac{1}{1+\chi\overline{n}} \left[
1-\frac{\chi(\frac{1}{2}+n -\overline{n})}
{1+\chi\overline{n}} \right].
\label{17}
\end{equation}
Because of the condition (\ref{10}) the above approximation
is justified for all terms of the sum in the
nominator of Eq. (\ref{16}),
since in these terms $|n-\overline{n}|$ is of the order of
magnitude of $\sqrt{\Delta n^2}$ or smaller,
and since   $\chi/(1+\chi \overline{n}) < 1/\overline{n}$
for any value of $\chi$.
The second term in the square brackets of Eq. (\ref{17}) therefore
leads to a contribution to $b$ that small in comparison to the
time-independent contribution 
of the first term.
Replacing  $\sigma_{n-1,n}(t)$ by its initial value
$\sigma_{n-1,n}(0)=\sqrt{n}\overline{\rho}_{n,n}$ 
in this small contribution and
taking into account that $\sum_{n=0}^{\infty} n^2 \overline{\rho}_{n,n}
= \overline{n}^2 + \Delta n^2$, we obtain
after minor algebra a time-independent
expression for $b$. The latter can be substituted into
Eq. (\ref{12}) to yield the approximative expression
\begin{equation}
\Delta \omega = \gamma - \frac{r\chi}{2(1+\chi\overline{n})}
\left[ 1- \frac{\chi}{1+\chi\overline{n}}
\left(\frac{1}{2}+\frac{\Delta n^2}{\overline{n}} \right) \right]
\label{18}
\end{equation}
for the laser linewidth.

In the next step we eliminate the 
injection rate $r$ by making use of the
photon-number balance equation (\ref{9}). For the case of
the laser the latter takes the form
\begin{equation}
\frac{r\chi}{2} \sum_{n=0}^{\infty}\frac{n+1}{1+\chi(n+1)}
\overline{\rho}_{n,n}=\gamma(\overline{n} - n_b) .
\label{19}
\end{equation}
Here again Eqs. (\ref{14}) and (\ref{15}) have been used 
together with Eq. (\ref{13}),
and the index of summation has been changed appropriately.
We proceed by applying the same approximation 
scheme that led to Eq. (\ref{18}) and perform the expansion
\begin{equation}
\frac{1}{1+\chi(n+1)} \approx
\frac{1}{1+\chi\overline{n}} \left[
1-\frac{\chi(1+n -\overline{n})}
{1+\chi\overline{n}} \right].
\label{20}
\end{equation}
Using the condition (\ref{10}) and the
relation \cite{nb}
\begin{equation}
n_b\ll \overline{n},
\label{20a}
\end{equation}
we obtain from Eq. (\ref{19})
after simple transformations the approximation
\begin{equation}
\frac{r\chi}{2(1+\chi\overline{n})}
=\gamma\left[ 1+ \frac{\chi}{1+\chi\overline{n}}
\left(1+\frac{\Delta n^2}{\overline{n}} \right)
-\frac{1+n_b}{\overline{n}}  \right].
\label{21}
\end{equation}
When  only the first term in the square
brackets is kept, Eq. (\ref{21})
 yields the familiar relation
\begin{equation}
\overline{n} = 
    \frac{r}{2\gamma}-\frac{1}{\chi} =
\frac{1}{\chi}
    \left(\frac{\alpha}{\gamma}-1\right),
\label{22a}
\end{equation}
where we have introduced the linear gain $\alpha = r\chi/2$.
However, in order to study the quantum limit of the linewidth, 
also the terms that are small in comparison to the leading term
have to be taken into account. 
When we substitute the expression (\ref{21}) for the factor
in front of the square brackets in Eq. (\ref{18}), 
keeping only the terms of lowest order in $1/\overline{n}$
and $\chi/(1+\chi\overline{n})$, respectively,
the contributions containing the variance cancel
and we finally arrive at the laser linewidth
\begin{equation}
\Delta\omega=\frac{\gamma}{2\overline{n}}
\left(\frac{2+\chi\overline{n}}{1+\chi\overline{n}}+2n_b\right).
\label{22}
\end{equation}
The quantum origin of the intrinsic laser linewidth is
revealed by noticing that along our lines we would have 
obtained the result 
$\Delta \omega = 0$  if the terms of the order $1/n$ arising 
from the commutation relation for the field operators had been 
neglected in Eqs. (\ref{16}) and (\ref{19}).
The two limiting cases $\chi\overline{n} \gg 1$ and
$\chi\overline{n} \ll 1$ correspond to a laser that is
operated far above threshold or near threshold, respectively,
as can be seen from Eq. (\ref{22a}).
Therefore it is apparent that the linewidth becomes inversely 
proportional to the mean photon number $\overline{n}$ only near 
threshold and far above threshold. In the intermediate regions 
the dependence on $\overline{n}$ is more complicated.

We note that because of the relation (\ref{22a}) the linewidth
can be expressed in terms of any three of the four
parameters $\overline{n}, \chi, \gamma$, and $\alpha$ (or $r$,
respectively). By inserting Eq. (\ref{22a}) into Eq. (\ref{22}) 
and thus eliminating $\overline{n}$,
we obtain another useful expression for the linewidth, 
\begin{equation}
\Delta\omega=\frac{\gamma\chi}{2}\,\frac{\gamma}{\alpha-\gamma}
    \left(1+\frac{\gamma}{\alpha}+2n_b\right).
\label{24}
\end{equation}
Eq. (\ref{22}) describes the dependence of the linewidth of a 
given laser on the mean photon number, while 
Eq. (\ref{24}) describes the dependence on the gain or, 
alternatively, on the above-threshold ratio, defined simply as 
the normalized gain, $\alpha / \gamma$.
Since the linear gain is easily measurable, this latter equation 
is the most important of the possible expressions for the 
intrinsic laser linewidth from the point of view of experimental 
accessibility.

In order to facilitate later comparison with the
standard result delivered by the phase-diffusion model,
we eliminate $\chi$ from  the linewidth expression (\ref{22}).
Using Eq. (\ref{22a}) one more times, we arrive immediately at 
\begin{equation}
\Delta\omega=\frac{\alpha + \gamma}{2\overline{n}}
    \frac{\gamma}{\alpha} + \frac{\gamma}{\overline{n}} n_{b} .
\label{23}
\end{equation}
Although it might seem from this expression that $\Delta \omega$
is proportional to $\overline{n}^{-1}$ in the entire region
of laser operation, this is not true since the gain $\alpha$
is not constant for a given laser but depends on the pumping rate
and is connected with $\overline{n}$ via the relation (\ref{22a}).
In contrast to this, the cavity decay constant $\gamma$ and the
saturation parameter $\chi$ are fixed, $\chi^{-1}$ being the 
saturation photon number for the lasing transition
between the atomic energy levels.

It is interesting to consider two important limiting cases.
In the far-above-threshold limit, where $\chi\overline{n} \gg 1$
or $\alpha \gg \gamma$, respectively,
Eqs. (\ref{22}) - (\ref{23}) yield the limiting value
\begin{equation}
 \Delta\omega_{\rm lim}
=\frac{\gamma}{2\overline{n}}(1+2n_b)
=\frac{\chi\gamma^2}{2\alpha}(1+2n_b).        
\label{24a}
\end{equation}
On the other hand, in the vicinity of the threshold,
where $0 < \alpha/\gamma -1 \, \ll \, 1$ \cite{footnote},
the linewidth depends on $\overline{n}$ in a different way,
described by $ \Delta\omega_{\rm thr}
=\frac{\gamma}{\overline{n}}(1+n_b)$.
This difference can be interpreted to be
due to the fact that the contribution
of intensity fluctuations to the linewidth increases when
the above-threshold ratio decreases, as we shall show
next.

In a quantized description of the radiation field,
intensity fluctuations are revealed in an enhancement of the 
photon-number variance as compared to the Poissonian value of 
$\Delta n^2 = \overline{n}$ that corresponds to a
constant intensity. Therefore, we first calculate
the steady-state photon-number variance $\Delta n^2$
of the laser, taking into account the
presence of thermal photons. To do so, we apply an approximation
method that has been developed previously by one of the authors
in order to investigate the photon statistics in saturated
multi-photon atom-field interaction \cite{m,h1}.
We start from the steady-state equation
\begin{equation}
\frac{d}{dt}(\Delta n^2) =
 \sum_{n=0}^{\infty} (n^2-2\overline{n}n)
 \dot{\overline{\rho}}_{n,n}=0 ,
\label{23a}
\end{equation}
and use the right-hand side of the master equation (\ref{5}),
together with Eqs. (\ref{2}) and (\ref{13}), 
in order to express $\dot{\overline{\rho}}_{n,n}$.
By taking into account Eqs. (\ref{14}) and (\ref{15})
and by appropriately
changing the index of summation in the individual terms,
the resulting equation can be transformed to yield
\begin{eqnarray}
 \sum_{n=0}^{\infty} \overline{\rho}_{n,n}
 \left \{ \alpha\,
\frac{(n+1)(2n+1-2\overline{n})}{1+\chi(n+1)}
-2\overline{n}\gamma(n_b - n) \right.  \nonumber \\
  \left.  +\gamma[n_b (n+1)(2n+1) - (1+n_b)n(2n-1)]
  \right\} = 0 ,
\label{25}
\end{eqnarray}
with  $ \alpha =r\chi/2$.
The following treatment again relies on the assumption of a
strongly peaked photon-number distribution subject to the
conditions (\ref{10}) that imply that the
approximation (\ref{20}) is valid.
When the latter is applied in Eq. (\ref{25}), we obtain a
term that is proportional to
$\sum_{n=0}^{\infty} n^3 \overline{\rho}_{n,n}
=\overline{n^3}$. Applying the approximation
\begin{equation}
\overline{n^3} = \overline{[\overline{n}+
(n-\overline{n})]^3} \approx \overline{n}^3
+3 \overline{n} \Delta n^2 , 
\label{25a}
\end{equation}
which is justified because of the 
condition (\ref{10}), we arrive at the equation
\begin{equation}
\frac{\alpha}{1+\chi\overline{n}}
\left[ \frac{2\Delta n^2}{1+\chi\overline{n}} +\overline{n} \right]
=\gamma[2\Delta n^2-\overline{n}(1+2n_b)].
\label{26}
\end{equation}
Here again the relation $\overline{n^2} = {\overline{n}}^2 +
\Delta n^2$ has been taken into account, and small contributions have
been neglected.
Finally, we again make use of the lowest-order photon-number balance
equation $\alpha/(1+\chi\overline{n}) = \gamma$ (cf. Eq.(\ref{22a})).
Thus, from Eq. (\ref{26}), after simple transformations,
we arrive at the relative photon-number variance of the laser,
\begin{equation}
\frac {\Delta n^2}{\overline{n}}=
 \left(1+\frac{1}{\chi\overline{n}}\right) (1+n_b)
 = \frac{\alpha}{\alpha -\gamma}(1+n_b).
\label{27}
\end{equation}
Clearly, when the thermal photon number $n_b$ is not small
in comparison to 1, its influence
on the photon-number variance of the laser is
crucial even for $n_b \ll \overline{n}$, as is its influence
on the intrinsic laser linewidth.
For $n_b = 0$ Eq. (\ref{27}) corresponds to the
standard result that is known from the literature
\cite{sargent,sctext,orszag}.
We note that in general the relative strength of the 
intensity fluctuations is characterized by the normalized
quantity
\begin{equation}
\frac{\langle a^{\dag^2}a^2 \rangle-\langle a^\dag a\rangle^2}
{\langle a^\dag a\rangle^2}
  = \frac{\overline{n(n-1)}}{\overline{n}^2} -1 =
  \frac{Q}{\overline{n}}.
\label{27a}
\end{equation}
Here we introduced the Mandel $Q$-parameter
\begin{equation}
Q \equiv\frac{\Delta n^2}{\overline{n}}-1
  =n_b+\frac{1+n_b}{\chi\overline{n}}
  =\frac{\gamma + n_b\,\alpha}{\alpha - \gamma},
  \label{27b}
\end{equation}
where Eq. (\ref{27}) has been applied.
Since the conditions (\ref{10}) imply that $Q \ll
\overline{n}$,  the intensity
fluctuations characterized by Eq. (\ref{27a}) are
extremely small under these conditions, and they
vanish in the limit $\overline{n} \rightarrow \infty$.
Therefore intensity fluctuations can be considered to
be a true quantum effect
in the above-threshold regime of the laser. 
Nevertheless their influence on the laser linewidth
cannot be neglected, because the non-zero value of the 
latter itself is a small quantum effect only.

With the help of  Eq. (\ref{27b}) the linewidth
equations (\ref{22}) - (\ref{23}) can be finally cast
into the form
\begin{equation}
\Delta\omega = \frac{\gamma}{2\overline{n}}
\left(1 + \frac{Q}{1+Q}\right)(1+n_b).
\label{28}
\end{equation}
Obviously, close to the threshold, where
$\alpha/\gamma -1 \ll 1$ \cite{footnote} and
hence $Q \gg 1$,
the contribution of the intensity fluctuations
to the intrinsic laser linewidth is the
same as that of the phase fluctuations. In this case the linewidth 
is twice as large as it would be without intensity fluctuations,
i. e. for $Q=0$.
                           
\section{Discussion}

Before discussing the analytic results in more detail,  
it seems appropriate to say a few words about their range of 
validity, determined by the applicability of our basic
assumptions (\ref{10}).
Making use of Eq. (\ref{27}), the second of the
inequalities (\ref{10}) can be transformed to yield the condition
$[(1+n_b)\alpha/(\alpha-\gamma)]^{1/2} \ll \overline{n}^{1/2}$.
When the laser is operated e. g. 10\% above threshold, i. e. for
$\alpha/\gamma = 1.1$, this requirement can be assumed to be
fulfilled if
$\overline{n}\gtrsim 10^3$
which, because of Eq. (\ref{22a}), corresponds to
$\chi \lesssim 10^{-4}$.
At higher above-threshold ratios
our approximation is valid for even smaller mean photon numbers,
or larger values of $\chi$, respectively. In general,
because of Eqs. (\ref{22a}) and (\ref{27}),
we can combine the two inequalities
of the condition (\ref{10}) to yield the requirement
\begin{equation}
\sqrt{\chi(1+n_b)} \, \ll \,
\frac{\alpha}{\gamma}-1 \, \ll \, \frac{1}{\chi} ,
\label{29}
\end{equation}
which has to be fulfilled for the linewidth formulae 
(\ref{22}) - (\ref{23}) to be valid.
Obviously, if $\chi \approx 10^{-6}$ and $n_b \ll 1$
as in typical continuous-wave gas lasers, our results
are already approximately  valid when the laser is
operated only more than 1\% above threshold, i. e. for
$\alpha / \gamma  \gtrsim 1.01$.

The linewidth formulae (\ref{22}) - (\ref{23})
constitute the main result of this paper. 
They differ from the standard laser linewidth formula,
\begin{equation}
\Delta\omega_{\rm pd}=\frac{\gamma+\alpha}{4\overline{n}}
=\frac{\gamma\chi}{4}\frac{\alpha+\gamma}{\alpha-\gamma}
=\frac{\gamma\chi}{4}+\frac{\gamma}{2\overline{n}},
\label{31}
\end{equation}
which has been derived for $n_b =0$ in the so-called 
phase-diffusion model neglecting intensity fluctuations 
\cite{pd}.
Here, for the last two steps, we used Eq. (\ref{22a})
and the equivalent relation
$\alpha=\gamma(1+\chi\overline{n})$ in order to transform
the standard result. 

In Fig. 1 the linewidth is plotted for different operating 
regimes
of the laser.  We emphasize that the curves representing our 
result,
(\ref{24}), are in perfect agreement with the numerical results, 
found previously for the laser linewidth by computing
the two-time correlation function of the field \cite{lu1}. From 
a comparison of Eq. (\ref{31}) to Eq. (\ref{23}), it is 
obvious that the phase-diffusion result
is only a good approximation for the linewidth
when $\alpha/\gamma \approx 2$,
whereas it underestimates the linewidth closer to the threshold. 
Higher above threshold, on the other hand, the 
linewidth is overestimated by Eq. (\ref{31}) which, for 
$\overline{n} \rightarrow \infty$, 
yields the intensity-independent residual linewidth $\Delta 
\omega = \gamma\chi/4$, instead of $\Delta \omega = 0$,
to be expected in the classical limit.
Moreover, we conclude from Fig. 1 that the
linewidth can be approximated by $\Delta\omega_{\rm lim}$, 
given by Eq. (\ref{24a}), 
provided that $\alpha/\gamma \gtrsim 5$.
The difference between our
result and the standard one in the far-above-threshold
region shows that the phase-diffusion assumption of the 
standard treatment overestimates
the contribution of phase fluctuations to the linewidth
the more the higher above threshold the laser is
operated, i. e. the stronger the effect of the
nonlinearity stemming from the gain saturation.
On the other hand, since intensity fluctuations are neglected,
the standard treatment underestimates the linewidth
in the near threshold regime and corrections are necessary to 
incorporate the effect of the super-Poissonian photon statistics,
as has been emphasized already by Lu \cite{lu1}.
As the above-threshold ratio increases and the intensity becomes
more stabilized, the effect of this underestimation decreases.
Fig. 1 suggests that the effects of overestimating phase
fluctuations, on the one hand, and underestimating intensity
fluctuations, on the other, just compensate approximately
when the laser is operated around 100\% above threshold.
 
\begin{figure}
\centerline{\epsfig{file=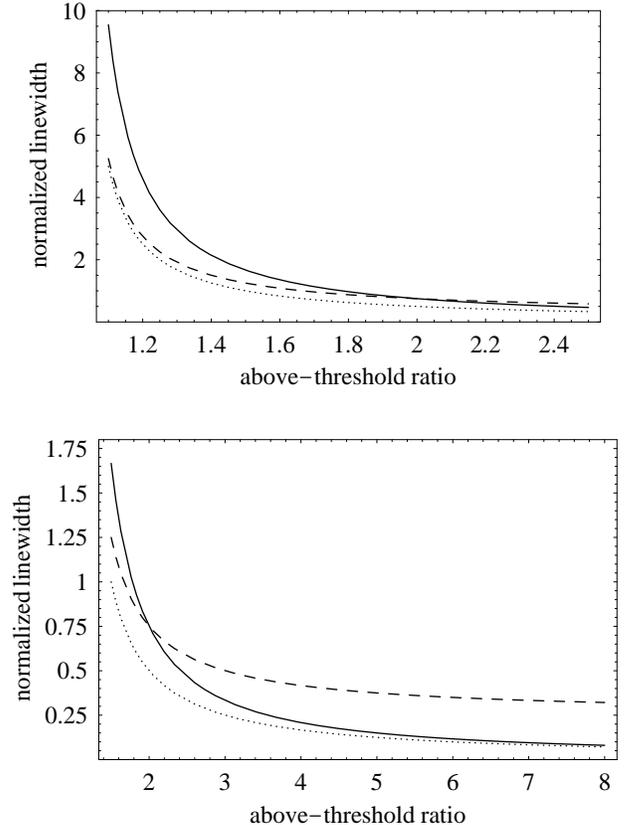,width=9cm}}
\caption                         
{Normalized laser linewidth $\Delta \omega /(\chi\gamma)$ versus 
the above-threshold ratio  $\alpha/\gamma$ for $n_b=0$. 
The full line corresponds to our formula, Eq. (\ref{24}).
For comparison, the curves resulting from the standard
expression $\Delta \omega_{\rm pd}$ (dashed line,  corresponding 
to Eq. (\ref{31})) 
and from the  approximation $\Delta \omega_{\rm lim}$
(dotted line, corresponding to Eq. (\ref{24a})) are 
also displayed.}
\end{figure}

In the following we shall discuss the reasons for the
discrepancy between our result and the standard one
in more detail. For this purpose, we first consider the 
different approximation 
methods that are employed in the literature for 
the derivation of Eq. (\ref{31}).   
In the most common approach, the master
equation of the density operator is transformed
into an equation for its $P$-representation.
After changing to polar coordinates by writing the complex field
amplitude as $ \epsilon = r\,{\rm exp}({\rm i}\phi)$,
the exact evolution equation for the quasi-probability
density $P(r,\phi)$ contains
derivatives with respect to $r$ and $\phi$ to all orders
\cite{sctext}. This is due to the nonlinearity of the
underlying master equation, which is revealed by the
denominators in Eqs. (\ref{14}) and (\ref{15}).
In the standard treatment it is assumed that
$P$ does not change along the radial coordinate,
corresponding to a neglect of intensity fluctuations, 
and that only derivatives up to second order
have to be taken into account. The latter assumption
is necessary to ensure phase diffusion, but it
becomes less and less justifiable as gain saturation,
hence nonlinearity
of the laser equations, increases.
In this case the simple phase diffusion model
can no longer be applied. In addition, 
mixed terms also become important in the 
evolution equation for $P$, which
are products of a differential operator acting on the 
amplitude $r$
and another one acting on the phase.  These, in turn, lead to 
cross correlations between the fluctuations of the 
intensity and the phase.

In the frame of the photon-number representation of the density
operator, the standard result has been derived
by applying the quantum fluctuation-regression theorem
and approximately investigating the decay of an initial value of
the field
instead of the two-time correlation function.
This is done  by means of determining the
lowest eigenvalue that characterizes a single decay
rate for all non-diagonal density-matrix elements $\rho_{n,n-1}$
\cite{sctext}.
In a more precise treatment a single decay rate
would have to be determined for the quantities
$\sqrt{n}\rho_{n,n-1}$ \cite{footnote1}, since the 
average field follows from
performing the sum over these quantities.
Moreover, with increasing above-threshold ratio the
influence of the nonlinearity also increases and
therefore the quantum fluctuation-regression theorem
cannot be applied anymore, in general \cite{regress}.
 
 With respect to the Heisenberg-Langevin 
approach, we mention that
a nonlinear $c$-number Langevin equation can be
derived for the complex field amplitude
$ \epsilon = r\,{\rm exp}({\rm i}\phi)$.
The coupling of the
fluctuations of the real amplitude $r$ and the phase $\phi$
is clearly obvious from this Langevin equation.
The derivation of the phase diffusion result, Eq. (\ref{31}), 
rests on implicitly making a factorization assumption
for expectation values containing the complex field amplitudes
in the denominator and their noise operators in the nominator
\cite{sctext}.
Because of the intensity fluctuations  
near threshold, and because of the nonlinearity due 
to gain saturation far above threshold, it would be
extremely difficult to go beyond this approximation.

We finally conclude that the laser linewidth
cannot be explained satisfactorily with the
help of the simple assumption that the
intensity is constant and the electric field phasor
executes a random walk in the complex plane as described by
phase diffusion. In the linear approximation, valid near threshold,
it is true that the behavior of the phase fluctuations alone
can be
described by phase diffusion, but intensity fluctuations contribute
to the linewidth as well.
Farther above threshold, on the other hand, the simple model
breaks down because the fluctuations of
the intensity and of the phase are coupled 
due to the nonlinearity of the gain, and the behavior of the
phase fluctuations cannot be characterized as simple
phase diffusion.
The frequently used procedure of
disregarding the microscopic intensity 
fluctuations and assuming that
the photon-number distribution is strictly Poissonian,
on the one hand, and not properly taking into account
or even completely neglecting gain saturation, on the other,
in general does not yield sufficiently
accurate results for the quantum-limited linewidth.

\section{Conclusions}

In this paper we have studied the quantum-limited
linewidth of a good-cavity laser
by determining the first-order correlation function of the
laser field at steady state. It is the decay of this correlation 
function and not
the phase fluctuations alone that determines coherence properties 
such as, e. g., the visibility 
of interference fringes. By taking the Fourier transform we obtained 
the power spectrum of the laser field and derived an analytical 
expression for the quantum-limited linewidth as a function of
the mean photon
number (see Eq. (\ref{22})) or of the above-threshold
ratio of the laser (see Eq. (\ref{24}). Our
analytical result is in perfect agreement with the results
of earlier numerical studies \cite{lu1}. We explicitly demonstrated
the effect of a super-Poissonian photon statistics (see Eq. (\ref{28}))
and showed that near threshold the linewidth
is considerably larger than the standard phase-diffusion result, 
(cf. Eq. (\ref{31})), where the intensity is assumed to be
constant, or the photon statistics to be Poissonian, respectively.
Although in most practical cases the laser linewidth is limited by 
the much larger technical noise and the intrinsic quantum limit 
cannot be reached,
there exists a variety of proposals to reach or even go beyond 
the quantum limit with the help of
sophisticated methods \cite{corr,wiseman}.
Our results show that for a precise quantum mechanical description 
of the laser linewidth it is necessary to directly calculate
the first-order correlation function of the laser field,
thus implicitly incorporating intensity fluctuations as well as phase
fluctuations, and to properly take into account the nonlinearity
of the gain.

 \section*{Acknowledgments}

J. B. wants to acknowledge the hospitality
of the Arbeitsgruppe ``Nichtklassische Strahlung''
at the Humboldt-University extended to him during his stay
in Berlin. The research
of J. B. was also supported by the Office of Naval
Research (Grant Number: N00014-92J-1233), by the 
Hungarian Science Research Fund (OTKA, Grant Number: T 030671) 
and by a grant from PSC-CUNY. 
U. H. is grateful for the hospitality
experienced during her visit at the
Janus Pannonius University in P\'{e}cs (Hungary).


\begin{thebibliography}{99}
\bibitem{nb} In fact, the validity of the relation
      $\overline{n} \gg n_b$ for the mean photon number of a laser
      above threshold is already implied in the second of the
      conditions (\ref{10}) since the latter would not be
      fulfilled for thermal radiation.
\bibitem{footnote}
     Note, however, that because of the assumption (\ref{10})
     our approximation procedure does not hold
    directly at the threshold, where the photon statistics does
    not differ appreciably from a thermal one and the photon number
    variance is still in the order of $\overline{n}^2$.
    (See also the discussion in Sec. V.)
\bibitem{pd} The standard linewidth formula (\ref{31}) has been derived 
    using different methods (see \cite{sargent,sctext,orszag}).
    It has been stated in Refs. \cite{sargent} and \cite{orszag} 
    that the result refers to the near-threshold region
    where $\alpha \approx \gamma$.
\bibitem{footnote1}
     For $n_b = 0$
the correct far-above threshold limit $\Delta \omega_{\rm lim} =
\gamma/(2\overline{n})$ has
been alternatively  derived recently \cite{wiseman}
from the two-time correlation function
by applying a method which is an extension of the
standard density-matrix investigation \cite{sctext}.
The procedure in Ref. \cite{wiseman} consists in determining a
single decay rate for the quantities
$\sqrt{n}\rho_{n,n-1}$ rather than for the non-diagonal
density-matrix elements $\rho_{n,n-1}$. It is applied to
a so-called ideal standard laser master equation
which is equivalent to the master equation
used in our model provided that in the coefficients
(\ref{14}) and (\ref{15}) the limit $\chi n \gg 1$ is
performed. This idealized formal equation guarantees
that the photon-number distribution is strictly Poissonian.
\end{thebibliography}
\end{document}